\documentclass[aps,twocolumn,prl,showpacs]{revtex4}
\usepackage{graphicx}% include figures
\usepackage{amssymb} % AMS math symbol
\usepackage{url}     % handle url
\usepackage{bm}      % bold math
\begin{document}

$\newcommand{\ohm}{\ensuremath{\,\Omega}}$

\title{Competing Channels for Hot Electron Cooling in Graphene}
\author{Qiong Ma$^{1}$}
\author{Nathaniel M. Gabor$^{1}$}
\author{Trond I. Andersen$^{1}$}
\author{Nityan L. Nair$^{1}$}
\author{Kenji Watanabe$^{2}$}
\author{Takashi Taniguchi$^{2}$}
\author{Pablo Jarillo-Herrero$^{1}$}
\email{pjarillo@mit.edu}
\affiliation{$^{1}$Department of Physics, Massachusetts Institute of Technology, Cambridge, MA 02139 USA}
\affiliation{$^{2}$National Institute for Materials Science, Namiki 1-1, Tsukuba, Ibaraki 305-0044, Japan}
\date{\today}

\begin{abstract}
We report on temperature dependent photocurrent measurements of high-quality dual-gated monolayer graphene (MLG) p-n junction devices. A photothermoelectric (PTE) effect governs the photocurrent response in our devices, allowing us to track the hot electron temperature and probe hot electron cooling channels over a wide temperature range (4 K to 300 K). At high temperatures ($T>T^*$), we found that both the peak photocurrent and the hot spot size decreased with temperature, while at low temperatures ($T<T^*$), we found the opposite, namely that the peak photocurrent and the hot spot size increased with temperature. This non-monotonic temperature dependence can be understood as resulting from the competition between two hot electron cooling pathways: (a) (intrinsic) momentum-conserving normal collisions (NC) that dominates at low temperatures and (b) (extrinsic) disorder-assisted supercollisions (SC) that dominates at high temperatures. Gate control in our high quality samples allows us to resolve the two processes in the same device for the first time. The peak temperature $T^*$ depends on carrier density and disorder concentration, thus allowing for an unprecedented way of controlling graphene's photoresponse.
%a photothermoelectric (PTE) effect governs the photoresponse at both the p-n and graphene-metal (G-M) interfaces, featuring a photocurrent that is non-monotonic in temperature and peaks at an intermediate temperature $T^*$. Spatial photocurrent microscopy shows that the size of photocurrent hot spots also varies non-monotonically with temperature and is maximized at the same $T^*$. This allows us to experimentally identify the competition of two different hot electron cooling pathways for the first time, namely momentum-conserving normal collisions (NC) and disorder-assisted supercollisions (SC). Furthermore, the peak temperature $T^*$ depends on carrier density and impurity concentration, thus allowing for an unprecedented way of controlling graphene's photoresponse.
\end{abstract}

%\pacs{72.20.-i, 72.80.Vp, 73.20.Hb, 73.22.Pr}
% 72.20.-i = Transport processes in semiconductors and insulators
% 72.80.Vp = Graphene electronic transport
% 73.20.Hb = Impurities at surfaces and interfaces
% 73.22.Pr = Graphene electronic structure

\maketitle

%%%%%%%%%%%%%%%%%%%%%%%%%%%%%%%%%%%%%%%%%%%%%%%%%%%%%%%%%%%%%%%%%%%%%%%%%
%%%%%%%%%%%%%%%%%%%%%%%%%%%%%%%%%%%%%%%%%%%%%%%%%%%%%%%%%%%%%%%%%%%%%%%%%
Slow electron-lattice thermal equilibration is responsible for a plethora of new optoelectronic \cite{Xu_nanolett, Song_nl, Gabor_science, Sun_naturenano}, transport \cite{Songdrag_prl, Songdrag_nl}, and thermoelectronic \cite{Betz_naturephy, Cho_naturemat} phenomena in graphene. The wide temperature ranges (lattice temperature, 4 K to 300 K) and long spatial scales in which hot carriers proliferate make graphene an ideal candidate for electronic energy transduction and numerous applications. Central to these are the unusual electron-phonon scattering pathways that dominate the cooling channels of graphene \cite{MacDonald_prl, Song_prl, Betz_naturephy, Graham_naturephy, Tielrooij_naturephy, Shi_nl}.

%Graphene is a promising candidate for photodetection due to its broad band absorption and fast response time \cite{Nair_science, Xia_naturenano}, but its low absorption efficiency is a limiting factor. A possible solution that has been proven to dramatically enhance light-graphene coupling strength, is to create hybrid devices by integrating graphene with plasmonic structures \cite{Echtermeryer_naturecomm}, Fabry-P$\acute{e}$rot cavities \cite{Furchi_nl} or highly absorptive materials with effective charge transfer \cite{Koppens_naturenano, Roy_naturenano}. Another potential approach is to increase the quantum efficiency through electron-electron scattering processes, which converts a single excited carrier into multiple lower-energy carriers, instead of losing the energy to heat \cite{Winzer_nl, Tielrooij_naturephy, Polini_naturecomm}. We provide new insight on the role of operation (environment) temperature in further confining energy to the electronic subsystem, steming from the competition between different cooling mechanisms. In graphene, there are two main relaxation pathways for hot Fermi-Dirac distributed carriers. Standard electron-phonon interactions allow for cooling through momentum-conserving normal collisions (NC) \cite{MacDonald_prl}, while in the absence of single phonon emission, electronic cooling instead occurs through disorder-induced supercollisions (SC), a disorder-dependent process by which the electron is cooled via the emission of multiple phonons \cite{Betz_naturephy, Graham_naturephy, Song_prl}.

Unlike other materials, electron-lattice cooling at room temperature in graphene is dominated by an extrinsic three-body process \cite{Song_prl}. This occurs when acoustic phonon emission is assisted by disorder scattering, as is called supercollisions (SC). SC dominates over the intrinsic momentum-conserving emission of acoustic phonons (NC) for high temperatures. At low temperatures, the intrinsic process is expected to be dominant. However, the intrinsic NC process has never been experimentally observed before \cite{Graham_naturephy, Graham_nl}.

Here, we report on temperature dependent spatially resolved photocurrent measurements of high-quality MLG p-n junction devices. At the p-n interface, the PTE effect dominates the photocurrent generation \cite{Song_nl, Gabor_science, Sun_naturenano} and exhibits a non-monotonic temperature dependence. We demonstrate that both the magnitude and spatial extent of the photoresponse are highly enhanced at an intermediate temperature $T^*$, indicating the coexistence of momentum-conserving NC cooling and disorder-assisted SC mechanisms. NC (SC) cooling dominates below (above) $T^*$, which can be tuned by varying the charge and impurity densities. In addition, we observed that the photoresponse at the graphene-metal (G-M) interface is also dominated by the PTE effect with a similar temperature dependent behavior. Lastly, we show that the dramatic suppression of hot carrier cooling at $T^*$ allows for non-local control of hot carrier dynamics. This involves top gate modulation of the photocurrent arising from illuminating the (distant) G-M interface.

%%%%%%%%%%%%%%%%%%%%%%      Figure 1     %%%%%%%%%%%%%%%%%%%%%%%%%%%%%%%%%%%%%%
%%%%%%%%%%%%%%%%%%%%%%%%%%%%%%%%%%%%%%%%%%%%%%%%%%%%%%%%%%%%%%%%%%%%%%%%%%%%%%%
\begin{figure}
\begin{center}
\vspace{-0.8cm}
\includegraphics[width=3.3in]{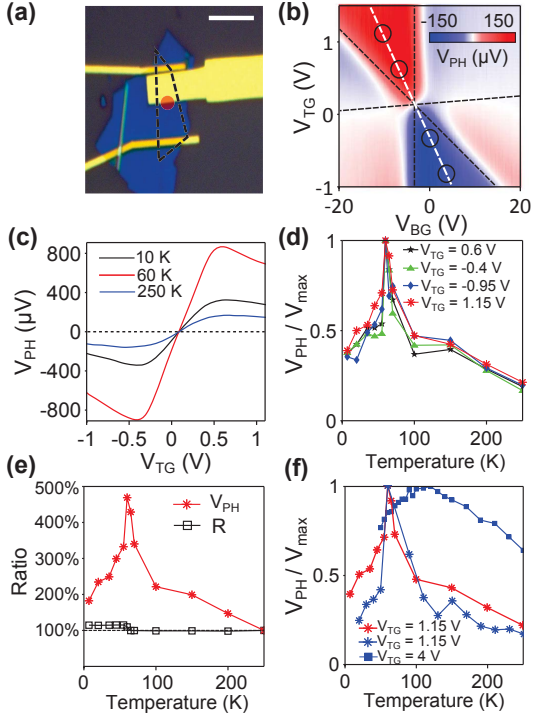}
\vspace{-0.7cm}
\caption{(a) Optical microscope image of the device incorporating boron nitride top-gate dielectric. The dashed black line marks the boundary of graphene underneath the boron nitride. (b) Photovoltage $V_\mathrm{PH}$ versus $V_\mathrm{BG}$ and $V_\mathrm{TG}$ at $T=250\,$K with laser fixed at the p-n interface (red dot in (a)). The white dashed line indicates $\mu_1=-\mu_2$. (c) Traces of (b) for $\mu_1=-\mu_2$ at different temperatures $T=10\,$K, $T=60\,$K, $T=250\,$K. Along $\mu_1=-\mu_2$, the cooling profile is symmetric from the p-n interface. (d) $V_\mathrm{PH}$ as a function of temperature (normalized to the maximum value) at particular points (black circles in (b)) along the $\mu_1=-\mu_2$ line. (e) Comparision between the temperature dependence of the photovoltage $V_\mathrm{PH}$ and the resistance $R$. Both $V_\mathrm{PH}$ and $R$ are normalized by their values at $T=250\,$K. (f) The temperature dependence of $V_\mathrm{PH}$ at low and high densities, showing a shift of $T^*$. Red curve: Device 1; blue curve: Device 2.} \label{F:tempdep}
\end{center}
\vspace{-1cm}
%\vspace{-0.5cm}
\end{figure}
%%%%%%%%%%%%%%%%%%%%%%%%%%%%%%%%%%%%%%%%%%%%%%%%%%%%%%%%%%%%%%%%%%%%%%%%%%%%%%%
%%%%%%%%%%%%%%%%%%%%%%%%%%%%%%%%%%%%%%%%%%%%%%%%%%%%%%%%%%%%%%%%%%%%%%%%%%%%%%%

Our MLG p-n junction devices are fabricated by micromechanical exfoliation, followed by standard e-beam lithography techniques to define contacts (Fig. 1a). Hexagonal boron nitride flakes ($10\,$-$20\,$nm thick) are then placed onto the samples as dielectric insulators by using a PMMA-transfer method, after which local top gates are fabricated to form p-n junctions in the center of the devices. In our experiments, the samples are kept in a liquid helium flow cryostat with an embedded resistive heater to give a precise temperature control from 4 K to above room temperature. We have measured 8 $\mathrm{SiO}_2$-supported exfoliated MLG devices with a high mobility of $\sim10,000\,\mathrm{cm}^2/\mathrm{Vs}$, all of which show similar results. The data presented in this paper was collected from two of them: Device 1 (8 $\mu$m long) and Device 2 (6 $\mu$m long).

By tuning back gate ($V_{\rm BG}$) and top gate ($V_{\rm TG}$) voltages independently, the junction can be operated in four different charge configurations: p-p, n-n, p-n and n-p \cite{Williams_science, PJH_prl2007, GHG_prl2007, Gabor_science, Sun_naturenano}. The photovoltage $V_\mathrm{PH}$ measured with the laser fixed at the p-n interface as a function of $V_\mathrm{BG}$ and $V_\mathrm{TG}$ exhibits six regions of alternating signs (Fig. 1b), which has been shown to be the fingerprint of the PTE effect \cite{Song_nl, Gabor_science, Sun_naturenano}. This six-fold pattern is observed over a wide range of lattice (environment) temperatures from $4\,$K up to $300\,$K. Fig. 1c shows slices of the photovoltage plot for $\mu_1=-\mu_2$ (dashed white line in Fig. 1b) at three representative temperatures ($10\,$K, $60\,$K and $250\,$K), where $\mu_1$ and $\mu_2$ are the chemical potentials in the single- and dual-gated regions, respectively. All three slices exhibit similar qualitative dependence on charge density, but the slice representing the intermediate temperature ($60\,$K) is the greatest in magnitude, indicating a non-monotonic dependence on temperature.

A detailed investigation of the relationship between photovoltage and lattice temperature is shown in Fig. 1d, where $V_\mathrm{PH}$ is plotted as a function of temperature at four different points (circles) along the $\mu_1=-\mu_2$ slice in Fig. 1b, all of which exhibit a dramatic enhancement at $T^*=60\,$K. In comparison, only minor differences are observed for the measured resistance, $R$, within the same temperature range. We plot $V_\mathrm{PH}$ (same as the red line in Fig. 1d) and $R$ (near the Dirac point) as a function of temperature in the same graph (Fig. 1e). Both $V_\mathrm{PH}$ and $R$ are normalized by their lowest values, which occur at $T=250\,$K. While $V_\mathrm{PH}$ shows an increase by as much as $500\%$ at $T^*$, $R$ stays fairly constant over the full temperature range (similar for $R$ measured away from the Dirac point). In Fig. 1f, we show the temperature dependence of $V_\mathrm{PH}$ collected from Device 2 at low (blue star) and high (blue square) densities, both of which exhibit non-monotonic behaviors but with an upwards shift of $T^*$ from low to high density.

This non-monotonic temperature dependence of the photoresponse is closely related to hot electron dynamics in graphene. To illustrate this, we begin by describing the photovoltage as $V_\mathrm{PH}=(S_1-S_2)\Delta T_\mathrm{pn}$ which determines the open-circuit PTE voltage generated from a sharply defined p-n junction \cite{Song_nl,Gabor_science} ($\Delta T_\mathrm{pn}$ is the electronic temperature increase at the p-n interface). We assume a linear response regime where $T_\mathrm{e} \gtrsim T_0$ \cite{Gabor_science} ($T_0$ is the lattice temperature) in the following analysis, which is consistent with the fact that all the measurements are performed in the linear power regime. By neglecting the temperature dependence of the resistance, $S_1-S_2$ is linear in $T$ \cite{Zuev_prl, Wei_prl}, indicating a strong temperature dependence embedded in $\Delta T_\mathrm{pn}$.

%%%%%%%%%%%%%%%%%%%%%%      Figure 2     %%%%%%%%%%%%%%%%%%%%%%%%%%%%%%%%%%%%%%
%%%%%%%%%%%%%%%%%%%%%%%%%%%%%%%%%%%%%%%%%%%%%%%%%%%%%%%%%%%%%%%%%%%%%%%%%%%%%%%
\begin{figure}
\begin{center}
\vspace{-0.1cm}
\includegraphics[width=3.5in]{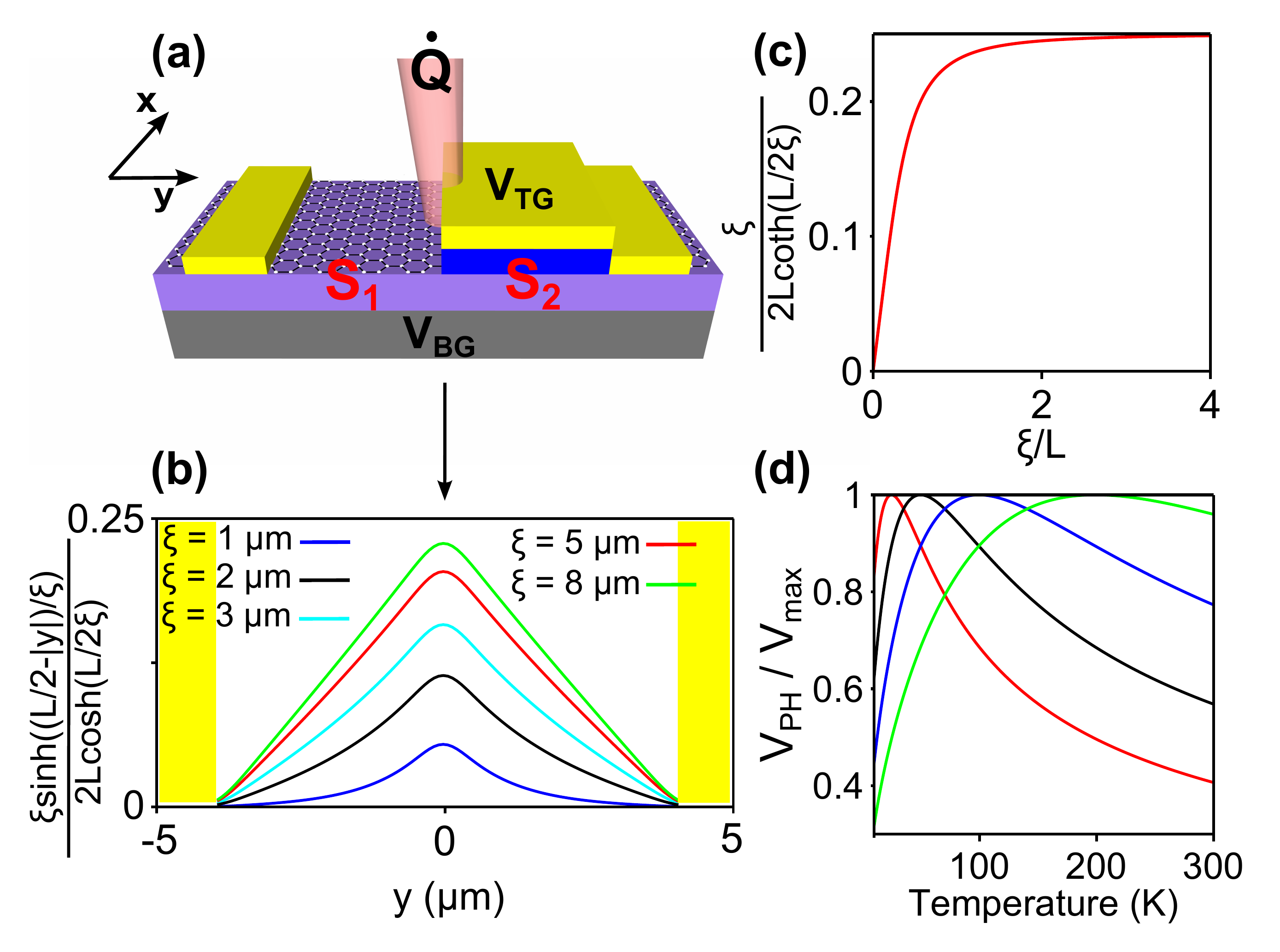}
\vspace{-0.7cm}
\caption{(a) Schematic of a dual-gated graphene p-n junction device (global backgate $V_\mathrm{BG}$ and local topgate $V_\mathrm{TG}$). $\dot{Q}$ denotes the rate at which heat enters the system by shining laser at the p-n interface. (b) Calculation of the quantity $\frac{\xi\sinh((\frac{1}{2}L-|y|)/\xi)}{2L\cosh(L/2\xi)}$, which is proportional to $T_\mathrm{e}-T_0$, as a function of $y$ for different cooling lengths. (c) Calculation of the quantity $\frac{\xi\sinh(L/2\xi)}{2L\cosh(L/2\xi)}$, which is proportional to the open-circuit photovoltage, as a function of $\xi/L$. $\xi$: cooling length. $L$: device length. (d) Calculation of the temperature dependence of photovoltage generated with the laser fixed at the p-n interface with varying $T_\mathrm{BG}$ by using $k_\mathrm{F}\ell=40$ (normalized to each peak). Red: $T_\mathrm{BG}=6\,$K; black: $T_\mathrm{BG}=12\,$K; blue: $T_\mathrm{BG}=24\,$K and green: $T_\mathrm{BG}=48\,$K for chemical potential $\mu=50, 100, 200, 400\,$ meV respectively.} \label{F:pnjunction}
\end{center}
\vspace{-1cm}
%\vspace{-0.5cm}
\end{figure}
%%%%%%%%%%%%%%%%%%%%%%%%%%%%%%%%%%%%%%%%%%%%%%%%%%%%%%%%%%%%%%%%%%%%%%%%%%%%%%%
%%%%%%%%%%%%%%%%%%%%%%%%%%%%%%%%%%%%%%%%%%%%%%%%%%%%%%%%%%%%%%%%%%%%%%%%%%%%%%%

With the laser focused on the p-n interface, a steady-state spatial profile of the electronic temperature $T_\mathrm{e}$ is established (Fig. 2a and 2b). Material parameters that can affect the profile include the thermal conductivity $\kappa$, the electronic specific heat $C_\mathrm{e}$, and the electron-lattice cooling rate $\gamma$. The combination of these three parameters generates a characteristic cooling length $\xi=(\kappa/\gamma C_\mathrm{e})^{\frac{1}{2}}$ for hot carrier propagation in the system \cite{Song_nl}. Due to the linear temperature dependence of both $\kappa$ (Wiedemann-Franz law) and $C_\mathrm{e}$, the temperature dependence of $\xi$ is embedded in $\gamma$. The analytical solution to the heat equation of the system is $T_\mathrm{e}(y)-T_0=\frac{\xi\sinh((\frac{1}{2}L-|y|)/\xi)}{2\cosh(L/2\xi)}\frac{\dot{Q}}{\kappa}$ (See Supplementary Information), where $\dot{Q}$ is the rate at which heat enters the system and $L$ is the device length ($\Delta T_\mathrm{pn}$ is the value at $y=0$). Fig. 2b shows the spatial profile of $T_\mathrm{e}-T_0$ in units of $\dot{Q}L/\kappa$, i.e., the dimensionless quantity $\frac{\xi\sinh((\frac{1}{2}L-|y|)/\xi)}{2L\cosh(L/2\xi)}$, for different values of $\xi$. The linear temperature dependence of $S$ cancels out that of $\kappa$ in the denominator of $\Delta T_\mathrm{pn}$ when we multiply these to find the photoresponse $V_\mathrm{PH}$. Consequently, the whole temperature dependence of $V_\mathrm{PH}$ is through $\xi$ and thus ultimately via the cooling rate $\gamma$ only. As can be seen from Fig. 2c, the photoresponse, which is proportional to $\frac{\xi\sinh(L/2\xi)}{2L\cosh(L/2\xi)}$, grows quickly with $\xi$ ($\gamma^{-\frac{1}{2}}$) and becomes saturated when $\xi$ approaches the system length $L$.

In order to understand the temperature dependence of $\gamma$, we consider possible hot carrier cooling pathways in graphene.  After initial relaxation of photo-excited carriers due to electron-electron scattering and optical phonon emission, the hot carrier distribution cools by emitting acoustic phonons \cite{MacDonald_prl}. A relevant cooling process is single acoustic phonon emission (NC), which gives a slow cooling rate $\gamma_\mathrm{NC}=A/T$ with a prefactor $A$ related to the charge density \cite{MacDonald_prl}. However, the disorder-assisted SC cooling gives rise to a competing cooling channel with a different cooling rate $\gamma_{\rm SC}=BT$, where the prefactor $B$ is related to the amount of disorder and the charge density \cite{Song_prl}. Therefore, NC and SC dominate at low and high temperatures respectively, with a cross-over temperature $T^*\approx(0.43\cdot k_\mathrm{F}\ell)^{\frac{1}{2}}T_\mathrm{BG}$ \cite{Song_prl}, where $T_\mathrm{BG}$ is the Bloch-Gr$\ddot\mathrm{u}$neisen temperature, and $k_\mathrm{F}\ell$ is the disorder-dependent mean free path as modeled in Ref. \cite{Song_prl}.

We see from the above equation that the optimal temperature for photodetection can be tuned by the disorder concentration (via $k_\mathrm{F}\ell$) and the charge carrier density (via $T_\mathrm{BG}$), which can be understood in terms of the relative weight of NC and SC cooling pathways. The available phase space for NC cooling is expanded when increasing the charge carrier density, and the SC channel is suppressed by reducing the disorder concentration. Both changes result in an increase in the temperature range dominated by NC and a decrease in the range of SC behavior, which will cause $T^*$ to increase. On the other hand, decreasing the carrier density and increasing the disorder amount will cause $T^*$ to shift to lower temperatures. The shifting of $T^*$ due to carrier density change is shown in Fig. 1f and Fig. 2d (simulation). To further verify the disorder relation, in-situ and systematic control of the disorder concentration is required. More detailed work needs to be done to uncover what type of disorder is dominant in assisting hot electron cooling. In any case, we want to emphasize that less disorder is always preferable in terms of the absolute efficiency of photocurrent at any temperatures.

The sensitivity of $T^*$ to the charge density and disorder concentration may account for the different (monotonic) temperature dependent behavior observed in previous studies \cite{Dawlaty_nl, George_nl, Plochocka_prb, Xu_nanolett, Strait_nanolett, Sun_naturenano, Winnerl_prl, Graham_naturephy, Graham_nl}. In addition, all the above arguments are based on the $T_\mathrm{e} \gtrsim T_\mathrm{0}$ condition, while otherwise we need to consider the full expression of the relative cooling weight between the SC and NC, which is derived in Ref. \cite{Song_prl} as $\frac{0.77}{k_\mathrm{F}\ell}\frac{T_\mathrm{e}^2+T_\mathrm{e}T_\mathrm{0}+T_{0}^2}{T_\mathrm{BG}^2}$. The overheating of electrons ($T_\mathrm{e}\gg T_\mathrm{0}$) will strongly enhance the SC weight even at low lattice temperatures, completely masking the NC processes.

%The out of equilibrium is critical for observing this. If the lattice and electron are in equilibrium, the cooling %coupling is essential, otherwise the large heat capacity strongly suppress this and also the non-monotonic %temperature dependence of the phonon thermal conductivity will yield a completely reverse trench for the %photoresponse.

The non-monotonic temperature dependence of hot electron cooling is reflected not only in the magnitude of the photocurrent, but also in its spatial profile. Fig. 3a shows the spatially resolved photocurrent microscopy ($V_\mathrm{SD}=0$) of Device 2. A strong photocurrent signal is observed at the p-n interface while the contact signals are strongly suppressed, which allows for independent extraction of the p-n signal profile. This signal decays with distance away from the p-n edge (denoted by a dashed black arrow in Fig. 3a) at different rates depending on temperature (Fig. 3b). The lowest decay rate is observed at the peak temperature $T^*=60\,$K, corresponding to the longest cooling length (theoretical simulations in Fig. 3b inset).

%%%%%%%%%%%%%%%%%%%%%%      Figure 3     %%%%%%%%%%%%%%%%%%%%%%%%%%%%%%%%%%%%%%
%%%%%%%%%%%%%%%%%%%%%%%%%%%%%%%%%%%%%%%%%%%%%%%%%%%%%%%%%%%%%%%%%%%%%%%%%%%%%%%
\begin{figure}
\begin{center}
\vspace{-0.7cm}
\includegraphics[width=2.5in]{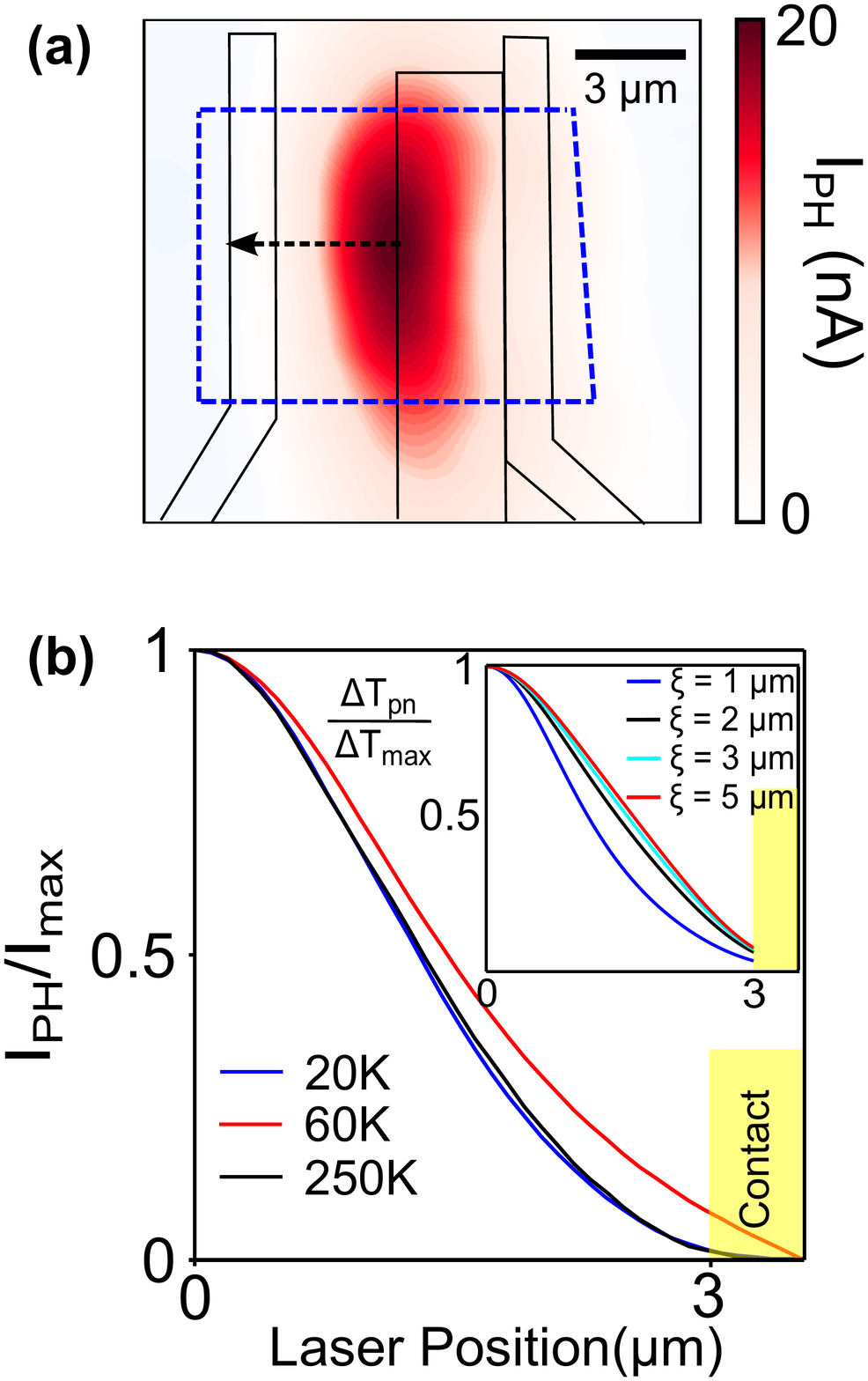}
\vspace{-0.2cm}
\caption{(a) Spatially resolved photocurrent map at $T=40\,$K for Device 2 ($V_\mathrm{SD}=0\,$ V, $V_\mathrm{BG}=25\,$ V, $V_\mathrm{TG}=-2\,$ V). Solid black lines mark the location of gold contacts and top gate electrode. Dashed blue lines mark the boundaries of graphene. (b) Photocurrent line traces (normalized to the peak) taken along the dashed black arrow in (a) at different temperatures. Laser position = 0 corresponds to the edge of the top-gate electrode. Inset: calculated electronic temperature increase at the p-n interface (normalized to the peak) as a function of the laser position along the dashed black arrow in (a) with varying cooling lengths. Note: both the measured and calculated spatial profiles here are (related to) the electronic temperature increase at the p-n interface as a function of the laser position while the profile in Fig. 2b is the electronic temperature increase as a function of the sample position $y$ with the laser fixed at the p-n interface.  } \label{F:cooling length}
\end{center}
\vspace{-1cm}
%\vspace{-0.5cm}
\end{figure}
%%%%%%%%%%%%%%%%%%%%%%%%%%%   Figure 3   %%%%%%%%%%%%%%%%%%%%%%%%%%%%%%%%%%%%%%
%%%%%%%%%%%%%%%%%%%%%%%%%%%%%%%%%%%%%%%%%%%%%%%%%%%%%%%%%%%%%%%%%%%%%%%%%%%%%%%

%%%%%%%%%%%%%%%%%%%%%%      Figure 4     %%%%%%%%%%%%%%%%%%%%%%%%%%%%%%%%%%%%%%
%%%%%%%%%%%%%%%%%%%%%%%%%%%%%%%%%%%%%%%%%%%%%%%%%%%%%%%%%%%%%%%%%%%%%%%%%%%%%%%
\begin{figure}
\begin{center}
\vspace{-1cm}
\includegraphics[width=3.4in]{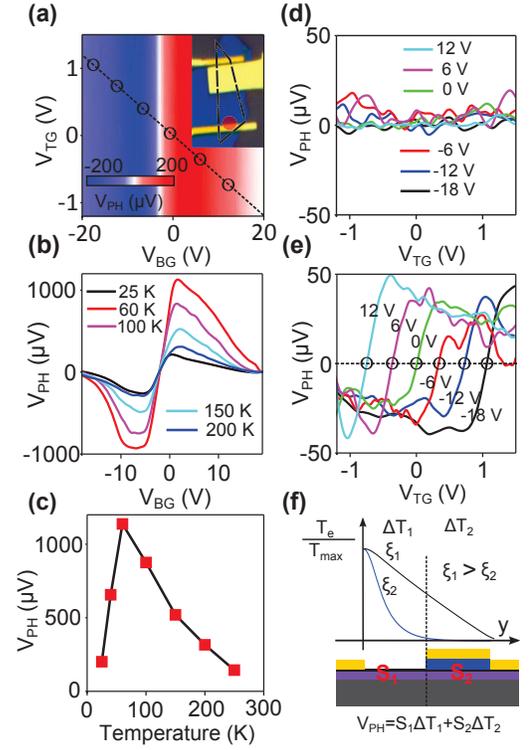}
\vspace{-1.1cm}
\caption{(a) $V_\mathrm{PH}$ as a function of $V_\mathrm{BG}$ and $V_\mathrm{TG}$ with the laser fixed at the contact away from the top gate (ret dot in the inset figure) at $T=250\,$K. Along the dashed black line, the dual-gated region is charge neutral. Inset: optical image of the device marking the laser position as a red dot fixed at the contact away from the top gate electrode.  (b) $V_\mathrm{PH}$ as a function of $V_\mathrm{BG}$ at a fixed top gate voltage $V_\mathrm{TG}=0.15\,$V for different lattice temperatures. (c) The maximum $V_\mathrm{PH}$ in (b) as a function of the lattice temperature. (d) $V_\mathrm{PH}$ as a function of $V_\mathrm{TG}$ at fixed back gates ($T=250\,$K, Vertical slices of (a)) exhibits no obvious top gate modulation. For each curve we have removed a constant background, i.e., the value along the diagonal dashed line in (a) (circles). (e) The same plot as in (d) for the peak temperature $T^*=60\,$K shows appreciable top gate modulation, which mimics the shape of the gate modulation of $S_2$. (f) Schematic of $T_\mathrm{e}$ as a function of the sample position $y$ with the laser fixed at the left contact for both long and short cooling length scenarios. Note the different formula for $V_{\rm PH}$ because laser is now at one contact instead of at the p-n junction.} \label{F:contact}
\end{center}
\vspace{-1.0cm}
%\vspace{-0.5cm}
\end{figure}
%%%%%%%%%%%%%%%%%%%%%%%%%%    Figure 4     %%%%%%%%%%%%%%%%%%%%%%%%%%%%%%%%%%%%
%%%%%%%%%%%%%%%%%%%%%%%%%%%%%%%%%%%%%%%%%%%%%%%%%%%%%%%%%%%%%%%%%%%%%%%%%%%%%%%

We now turn to the G-M interface of Device 1, where we also see evidence of a PTE response to laser illumination. In order to avoid ambiguity, we fixed the laser at the interface away from the top gate electrode. Fig. 4a shows the photovoltage $V_\mathrm{PH}$ as a function of $V_\mathrm{BG}$ and $V_\mathrm{TG}$ at 250 K, exhibiting complete reversal of polarity with respect to $V_\mathrm{BG}$, which is consistent with previous studies \cite{Lee_nnano, Park_nanolett, Mueller_prb, Xia_nl}. Fig. 4b plots $V_\mathrm{PH}$ slices as a function of $V_\mathrm{BG}$ (fixed $V_\mathrm{TG}$) at various temperatures, showing once again that the photovoltage is maximized at an intermediate temperature (60 K) and thus has a non-monotonic temperature dependence. The full temperature dependent behavior of $V_\mathrm{PH}$ is shown in Fig. 4c, where the maximum values of $V_\mathrm{PH}$ are plotted against temperature. We emphasize that this non-monotonic temperature dependence due to the hot electron cooling is unique to the PTE response and is not expected from the conventional photovoltaic (PV) effect, in which the seperation of excited carriers by the built-in electric field leads to a net current \cite{Lee_nnano, Park_nanolett, Mueller_prb, Xia_nl}. Therefore, this serves as a strong indication that the PTE effect dominates the response to laser illumination of the G-M interface, consistent with recent reports where the photovoltaic contribution at 800 nm wavelength is relatively small \cite{Ferrari_arxiv2014}.

Another important fact that can be extracted from Fig. 4a is that $V_\mathrm{PH}$ exhibits very little dependence on $V_\mathrm{TG}$ at $T=250\,$K. This is shown more clearly in Fig. 4d, where $V_\mathrm{PH}$ is plotted as a function of $V_\mathrm{TG}$ at different back gate voltages (vertical slices of Fig. 4a).  Each curve has been subtracted by the value along the diagonal dashed line in Fig. 4a, which defines the charge neutrality point of the dual-gated region. Indeed, no obvious top gate dependence is observable other than random fluctuations. In striking contrast, the same plot as Fig. 4d, but at the peak temperature $T^*=60\,$K instead, exhibits clear top gate modulation (Fig. 4e), indicating nonlocal hot carrier transport enhanced by a long cooling length.

This is further illustrated in Fig. 4f. When the cooling length $\xi$ is short either due to NC at low temperature or SC at high temperature, the hot carriers strongly thermalize with the lattice before reaching the top-gated region. This results in a low temperature gradient $\Delta T_2$ and thus a low PTE voltage in that region. Therefore, it is difficult to observe the modulation of $S_2$ by the top gate. In contrast, at the peak temperature $T^*$, the energy loss from the electronic system to the lattice is minimized. Hot carriers feature long relaxation lifetimes and long spatial propagation, leading to a considerable $\Delta T_2$ to drive $S_2$. In this regime, the top gate modulation is readily observable (Fig. 4e).

In conclusion, we have observed a strong non-monotonic temperature dependent behavior of the PTE response of high quality graphene p-n and G-M junctions. This behavior originates from two competing mechanisms for hot carrier cooling. At the peak temperature, hot carriers cool the slowest, resulting in a 5-fold increase in the photocurrent generation with respect to room temperature and a dramatic nonlocal phenomenon. This optimal temperature for maximal hot carrier extraction is controllable by carrier density and disorder concentration, which may pave the way for the design of more efficient graphene hot carrier devices.

%%%% ending %%%%

We thank J. C. W. Song and L. S. Levitov for numerous fruitful discussions. This work has been supported by AFOSR grant number FA9550-11-1-0225 (measurement and data analysis) and a Packard Fellowship. Device fabrication was supported by a seed fund from S3TEC, an Energy Frontier Research Center funded by DOE, Office of Science, BES under Award number DE-SC0001299/DE-FG02-09ER46577. This work made use of the Materials Research Science and Engineering Center Shared Experimental Facilities supported by the National Science Foundation (NSF) (award no. DMR-0819762) and of Harvard$'$s Center for Nanoscale Systems, supported by the NSF (grant ECS-0335765).

\bibliography{Tempdep_V11}

\newpage

\end{document}